\documentclass[useAMS,usenatbib]{mn2e}

\usepackage[psamsfonts]{amssymb}
\usepackage[dvips]{graphicx}
\usepackage{amsmath,alltt}

\title[Stellar Populations in Globular Cluster Cores]{Stellar
  Populations in Globular Cluster Cores:  Evidence for a Peculiar
  Trend Among Red Giant Branch Stars}
\author[N. Leigh, A. Sills and C. Knigge]{N. Leigh$^{1}$,
  A. Sills$^{1}$\thanks{E-mail:
leighn@mcmaster.ca (NL); asills@mcmaster.ca (AS)} and C. Knigge$^{2}$\thanks{E-mail:
christian@astro.soton.ac.uk (CK)}
\\
$^{1}$Department of Physics and Astronomy, McMaster University, 
1280 Main St. W., Hamilton, ON, L8S 4M1, Canada\\
$^{2}$School of Physics and Astronomy, Southampton University,
  Highfield, Southampton, SO17 1BJ, UK} 
\begin{document}


\pagerange{\pageref{firstpage}--\pageref{lastpage}} \pubyear{2009}

\maketitle

\label{firstpage}

\begin{abstract}
We investigate the relationship between the mass of a
globular cluster core and the sizes of its various stellar 
populations in a sample of 56 globular
clusters.  The number of core red giant branch stars is found to scale
sub-linearly with core mass at the 3-$\sigma$ confidence
level, whereas the relation is linear to within one standard deviation
for main-sequence turn-off and sub-giant branch stars.  We interpret
our results as evidence for a surplus of red giant branch stars in the
least massive cluster cores which is not seen for main-sequence
turn-off and sub-giant branch stars.  We explore various possibilities 
for the source of this discrepancy, discussing our results primarily
in terms of the interplay between the cluster dynamics and stellar
evolution.
\end{abstract}


\section{Introduction} \label{intro}

Studying the radial distributions of the various stellar 
populations (red giant branch, horizontal branch, main-sequence,
etc.) found in globular clusters (GCs) can provide useful hints 
regarding their dynamical histories.  As clusters evolve, they are
expected to undergo relatively rapid mass stratification as a
consequence of two-body relaxation, with the heaviest stars quickly
sinking to the central cluster regions
\citep{spitzer87}.  The shorter the relaxation time (typically
evaluated at the half-mass radius), the quicker this process occurs.  
Clusters tend towards dissolution as two-body
relaxation progresses and they lose mass due to stellar evolution and
the preferential escape of low-mass stars.  External effects like
tidal perturbations, encounters with giant molecular clouds, and
passages through the Galactic disk serve only to speed up the process
\citep[e.g.][]{baumgardt03, kupper08}.  Stellar evolution complicates
this otherwise simple picture of GC evolution, however.  Stars are
expected to change in size and lose mass as they evolve, often
dramatically, and this could significantly impact the outcomes of
future dynamical interactions with other stars.  For instance, a
typical star in a GC is expected to expand by up to a few orders of
magnitude as it ascends the red giant branch (RGB) and will shed up to
a quarter of its mass upon evolving from the tip of the RGB to the 
horizontal-branch (HB) \citep[e.g.][]{caloi08, lee94}.  

Red giant branch stars have been reported to be deficient in the cores
of some Milky Way (MW) GCs.  For instance, \citet{bailyn94} found that the
morphology of the giant branch in the dense core of 47 Tuc differs
markedly from that in the cluster outskirts.  In particular, there
appear to be fewer bright RGB stars in the core as well as an
enhanced asymptotic giant branch (AGB) sequence.  While
a similar deficiency of bright giants has been observed in the cores
of the massive GCs NGC 2808 and NGC 2419, better agreement between the
observations and theoretical luminosity functions obtained with the
Victoria-Regina isochrones was found for M5 \citep{sandquist07,
  sandquist08}.  \citet{sandquist07} speculate that the giant star 
observations in NGC 2808 could be linked to its unusually blue horizontal
branch if a fraction of the stars near the tip of the RGB experience
sufficiently enhanced mass loss that they leave the RGB early.
Alternatively, \citet{beer03} suggest that RGB stars could be depleted
in dense stellar environments as a result of collisions between red
giants and binaries. 

While stellar populations have been studied and compared on an
individual cluster basis, a statistical analysis in which their core
populations are compared over a large sample of GCs is ideal for
isolating trends in their differences.  Though a handful of studies of this
nature have been performed \citep[e.g.][]{piotto04}, we present an
alternative method by which quantitative constraints can be found for
the relative sizes of different stellar populations.  Specifically, a 
cluster-to-cluster variation in the central stellar mass function can
be looked for by comparing the core masses to the sizes of their
various stellar populations.  Since stellar evolution is the principal
factor affecting their relative numbers in the core, we expect the
size of each stellar population to scale linearly with the core mass.
If not, this could be evidence
that other factors, such as stellar dynamics, are playing an
important role.  
In this paper, we present a comparison of the core 
RGB, main-sequence (MS) and HB populations of 56 GCs.
In particular, we use star 
counts for each stellar population to show that RGB stars are either 
over-abundant in the least massive cores or under-abundant in the most
massive cores, and that this effect is not
seen for MS stars.  We present the data in Section~\ref{data} and our
methodology and results in Section~\ref{results}.  In
Section~\ref{discussion}, we discuss the implications of our results
and explore various possibilities for the source of the observed
discrepancy between RGB and MS stars.  Concluding remarks are
presented in Section~\ref{summary}. 

\section{The Data} \label{data}

\begin{figure}
\begin{center}
\includegraphics[width=3.2in]{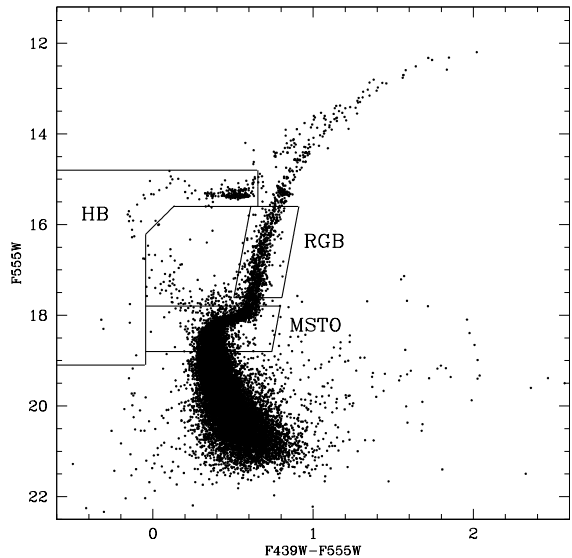}
\end{center}
 \caption{Colour-magnitude diagram for
  NGC 362 in the (F439W-F555W)-F555W plane.  Boundaries enclosing the
  selected RGB, HB and MSTO populations are shown.  
\label{fig:ngc0362_labels}}
\end{figure}

Colour-magnitude diagrams (CMDs) taken from \citet{piotto02} are used to
obtain star counts for the RGB, HB, MS and blue straggler (BS)
populations in the cores of 56 GCs.  We apply the same selection 
criterion as outlined in Leigh, Sills \& Knigge (2007) to derive our
sample as well as to define the location of the main-sequence turn-off
(MSTO) in the (F439W-F555W)-F555W plane.  An example of this selection
criterion, applied to the 
CMD of NGC 362, is shown in Figure \ref{fig:ngc0362_labels}.  We
include all stars in the \citet{piotto02} database.  Since
\citet{piotto02} took their HST snapshots with the centre of the PC
chip aligned with the cluster centre, a portion of the cluster core
was not sampled for most GCs.  We have therefore applied a geometric
correction factor to the star counts in these clusters in order to
obtain numbers that are representative of the entire core
\citep{leigh07, leigh08}.  The total number of stars in the core is
found by summing over all stars brighter than 1 mag below the
MSTO and then multiplying by the appropriate geometric correction factor.

Errors on the number of stars for each stellar population were
calculated using Poisson statistics.  Core radii,  
distance moduli, extinction corrections, central luminosity
densities and central surface brightnesses were taken from the Harris
Milky Way Globular Cluster catalogue \citep{harris96}.  Calibrated
apparent magnitudes in the F555W, F439W and Johnson V bands 
were taken from \citet{piotto02}.

\section{Results} \label{results}

This paper focuses on the core RGB, MS and HB populations of 56
GCs, comparing their numbers to the core masses.  Note that we
are focusing on the total number of stars in the core as a proxy for the
core mass instead of the total luminosity in the core in order to
avoid concerns regarding cluster-to-cluster variations in the central
stellar mass function and selection effects.  Given that a single
bright HB star can be as luminous as 100 regular MS stars, 
a small surplus of bright stars could have a dramatic impact on
the total luminosity.  
Therefore, the total number of stars in
the core is a more direct and reliable estimate for the core mass than
is the core luminosity.

Upon plotting the logarithm of the number of core RGB stars
versus the logarithm of the total number of stars in the core and
performing a weighted least-squares fit, we find a relation of
the form:
\begin{equation}
\label{eqn:power-laws}
\log (N_{RGB}) = (0.89 \pm 0.03)\log (N_{core}/10^3) + (2.04 \pm 0.02)
\end{equation}
The sub-linear slope is either indicative of a surplus of RGB stars in the
least massive cluster cores or a deficiency in the most massive cores.
Errors for lines of best fit were  
found using a bootstrap methodology in which we generated 1,000 fake
data sets by randomly sampling (with replacement) RGB counts from the
observations.  We obtained lines of best fit for each fake data set, fit a
Gaussian to the subsequent distribution and extracted its standard
deviation.  In order to avoid the additional uncertainty introduced
into our RGB number counts from trying to distinguish AGB stars from RGB
stars, as well as the difficulty in creating a selection criterion that
is consistent from cluster-to-cluster when including the brightest
portion of the RGB, stars that satisfy the RGB 
selection criterion shown in Figure \ref{fig:ngc0362_labels} are
referred to as RGB stars throughout this paper.  Note that it is
the brightest portion of the RGB that should be the most affected by
stellar evolution effects such as mass-loss.  If we
extend our selection criterion to include the entire RGB, however, our
results remain unchanged.

Interestingly, MS plus sub-giant branch stars (hereafter
collectively referred to as MSTO stars, the selection criterion for
which is shown in Figure 1) show a more linear relationship than do RGB
stars and appear to dominate the central star counts.  If we count
only those stars having a F555W mag within half a magnitude above and
below the turn-off, we obtain a relation of the form:
\begin{equation}
\log (N_{MSTO}) = (1.02 \pm 0.01)\log (N_{core}/10^3) + (2.66 \pm 0.01)
\end{equation}
A nearly identical fit is found when counting only those
stars having a F555W mag between the turn-off and one
magnitude fainter than the turn-off. 

We also tried plotting the logarithm of the number of core
helium-burning stars (labeled HB in Figure~\ref{fig:ngc0362_labels})
versus the logarithm of the number of stars in the core, yielding a
relation of the form:
\begin{equation}
\log (N_{HB}) = (0.91 \pm 0.10)\log (N_{core}/10^3) + (1.58 \pm 
0.05)
\end{equation}   
Note the large uncertainty associated with the fit, indicating that
the slope is consistent with both those of the RGB and MSTO samples.
We will discuss this stellar population further in
Section~\ref{discussion}.

\begin{figure}
\begin{center}
\includegraphics[width=3.2in]{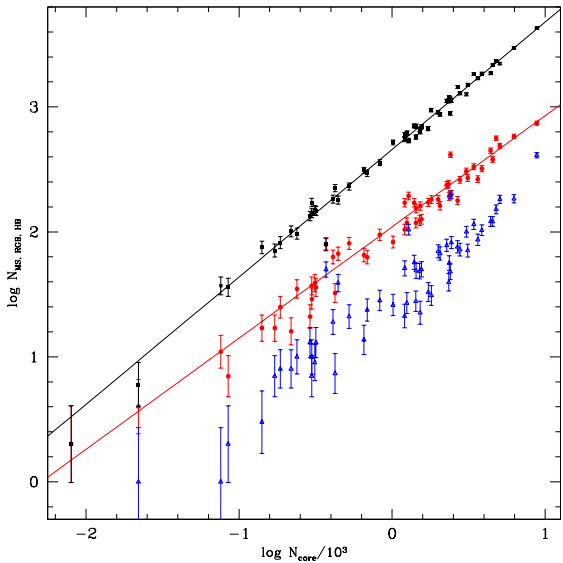}
\end{center}
\caption[N$_{core}$ versus N$_{MS,RGB,HB}$]{The number of RGB
  (red circles), MSTO (black squares) and HB (blue triangles) stars
  found within the cluster core plotted versus the total number of
  stars in the core brighter than 1 mag below the MSTO, along with the 
  corresponding lines of best fit for the RGB and MSTO samples.
\label{fig:N_vs_ncore}}
\end{figure}

The number of MSTO, RGB and HB stars are shown as a function of the
total number of stars in the core in Figure \ref{fig:N_vs_ncore}.
Interestingly, the blue stragglers in our sample also scale
sub-linearly with core mass, albeit more dramatically, obeying a
relation of the form N$_{BS}$ $\sim$ M$_{core}^{0.38 \pm 0.04}$
\citep{knigge09}.  Note that N$_{core}$ can be used interchangeably
with M$_{core}$.  In this case, we obtain a fit of:
\begin{equation}
\label{eqn:power-law-bs}
\log (N_{BS}) = (0.47 \pm 0.06)\log (N_{core}/10^3) + (1.22 \pm 0.02)
\end{equation}

In an effort to explore the influence of selection effects, we re-did our
plots having removed from our sample clusters denser than log $\rho$
$>$ 10$^5$ L$_{\odot}$ pc$^{-3}$ since we are the most likely to be
under-counting stars in the most crowded cluster cores where blending
of the stellar light is the most severe.  This cut also removes from
our sample the post-core collapse (PCC) clusters for which the core
radii are poorly defined since King models are known to provide a poor
fit to the observed surface brightness profiles in these clusters.
Similarly, we applied a cut in the central surface brightness,
removing from our sample clusters satisfying $\Sigma_0$ $<$ 15.1 V mag
arcsec$^{-2}$.  Finally, since clusters having both high surface
brightnesses and small cores are the most likely to suffer from
selection effects, we also tried adding to 
the aforementioned cut in $\Sigma_0$ a cut in the angular core radius,
removing clusters with r$_c$ $\le$ 0.05'.  In all cases, the
sub-linear power-law index reported for the RGB remains unchanged to
within one standard deviation of our original result.  Selection 
effects do not appear to be the source of the observed sub-linearity,
though it is clear that its effects must properly be accounted for in
future studies.

In order to assess the effects of age-related cluster-to-cluster
variations in the stellar mass function, as well as to test our
assumption that the number of stars in the core provides a reliable
estimate for the core mass, we have obtained MSTO 
masses for most of the GCs in our sample.  We fit theoretical
isochrones provided in \citet{pols98} to the cluster CMDs, using
the bluest point along the MS of a given isochrone as a proxy for the
MSTO mass.  Isochrones were calculated
using the metallicities of \citet{piotto02} and cluster ages were
taken from \citet{deangeli05} using the Zinn \& West (1984)
metallicity scale.  Core masses were estimated by multiplying the mass
corresponding to the MSTO (m$_{MSTO}$) by the number of stars in the
core brighter than 1 mag below the turn-off.  This is a reasonable
assumption given the very small dispersion in the ages of MW
GCs \citep{deangeli05} and the fact that we are only considering stars
brighter than 1 mag below the TO.  Consequently, the range of stellar
masses upon which we are basing our 
number counts is very small.  Our results remain entirely unchanged
upon using M$_{core}$ $\sim$ N$_{core}$m$_{MSTO}$ as a proxy for the
core mass instead of pure number counts.

In order to further check the sensitivity of our results to our estimate for
the core masses, we re-did all plots shown in
Figure~\ref{fig:N_vs_ncore} using various approximations for the total core
luminosity instead of pure number counts.  Core luminosities are
calculated in the Johnson V band directly from the stellar fluxes
which are summed over all stars in the core and then multiplied by the
appropriate geometric correction factor.  We also adopted L$_{core}$ =
$\frac{4}{3}\pi$r$_c^3$$\rho_0$, where $\rho_0$ is the
central luminosity density in L$_{\odot}$ pc$^{-3}$ taken from
\citet{harris96}.  Additionally, since the number of core RGB stars is
in reality a projected quantity, we tried plotting N$_{RGB}$ versus
L$_{core}$ = $\pi$r$_c^2$$\Sigma_0$, where 
$\Sigma_0$ is the central surface brightness in L$_{\odot}$ 
pc$^{-2}$, so that we are consistently comparing two projected
quantities.  Finally, we can adopt slightly more realistic
estimates for the total core luminosity by integrating over King
density profiles.  We fit single-mass King models calculated using the
method of \citet{sigurdsson95} to the surface brightness profiles of the
majority of the clusters in our sample using the concentration
parameters of \citet{mclaughlin05} and the central luminosity
densities of \citet{harris96}.  We then integrated the derived
luminosity density profiles numerically in order to estimate the total
stellar light contained within the core.  After removing clusters
labelled as post-core collapse in \citet{harris96} for which King
models are known to provide a poor fit, we once again compared the
integrated core luminosities to the number of RGB stars in the core.
For all four of these estimates for the total core luminosity, we
find that our fundamental results remain unchanged, with the power-law
index for RGB stars remaining sub-linear at the 3-$\sigma$ level.
Therefore, we conclude that our result is robust to changes in choices
of cluster and population parameters.  

\section{Discussion} \label{discussion}

We have shown that the number of RGB stars in globular cluster cores
does not directly trace the total stellar population in those cores.  In
particular, the number of RGB (but not MSTO) stars 
scales sub-linearly with core mass at the 3-$\sigma$ level.  Given
that the MS lifetime is expected to be a factor of 10-100 
longer than that of the RGB sample \citep{iben91}, the ratio
N$_{MSTO}$/N$_{RGB}$ indicates that the relative sizes of these stellar
populations are in better agreement with the expectations of stellar evolution
theory in the most massive cores.  This suggests that our results are
consistent with a surplus of RGB stars in the least massive cores.  We 
discuss below some of the key considerations in understanding the
evolution of GC cores and the stars that populate them in an effort to
explain our result.

\subsection{Stellar evolution}

Could this trend be a reflection of a stellar evolution process?  The
evolution and distribution of stellar populations can be thought 
of as the sum of many single stellar evolution tracks, which depend
only on a star's mass and composition.  Since there is no relation
between a cluster's mass and its metallicity \citep{harris96} and the
dispersion in the relative ages of MW GCs is quite small
\citep{deangeli05}, there is no reason to expect the RGB 
lifetime to depend on the cluster mass.  On the other hand, recent
studies suggest that 
the chemical self-enrichment of GCs during their early evolutionary
stages could help to explain many of the population differences
observed among them \citep[e.g.][]{caloi07}.  In particular, many of
the most massive GCs are thought to be enriched in helium and this is
expected to reduce the time scale for stellar evolution
\citep[e.g.][]{romano07}.  While this scenario predicts a deficiency
of RGB stars in the most massive cores, it would contribute to
depressing the slope of the RGB sample relative to that of the MSTO
population in Figure~\ref{fig:N_vs_ncore}.  

\subsection{Single star dynamics} \label{dynamics}

Two-body relaxation is the principal driving force behind the dynamical
evolution of present-day GCs, slowly steering them towards a state of
increased mass stratification as predominantly  
massive stars fall into the core and typically low-mass stars are
ejected via dynamical encounters.  The relaxation time increases
with the cluster mass \citep{spitzer87} and the variance in the relative ages
(and hence MSTO masses) of MW GCs is quite small
\citep{deangeli05}.  Therefore, it is the
least massive clusters that should show signs of being the
most dynamically evolved.  This assumes that cluster-to-cluster
variations in the initial mass function and the degree of
initial mass segregation are small.  In general, however,
proportionately fewer massive stars should have had sufficient time to
migrate into the most massive cores, while 
fewer low-mass stars should have been ejected out.  While qualitatively
correct, this effect should contribute little to the observed
difference between the core RGB and MSTO populations since RGB stars
are only slightly more massive \citep[e.g.][]{demarchi07}.

Stars expand considerably as they ascend the RGB.  Both the increase
in collisional cross-section and the change 
in the average stellar density could have an important bearing on the
outcomes of dynamical interactions involving RGB stars.  Indeed,
\citet{bailyn94} suggests that interactions between giants and other
cluster members in the core could strip the outer envelope of the
giant before it has a chance to fully ascend the RGB.  Since our
adopted RGB selection criterion does not include the brightest giants, 
we are only considering giants that are larger than MSTO stars
by a factor of $\sim$ 10 \citep{iben91}.  This small degree of
expansion will have only a minor effect on the collision rate.  Any
scenario that relies on dynamical encounters to explain a depletion of
RGB stars should be operating in very dense cores.  Our results are
consistent with some of the densest clusters in our sample having a
surplus of giants, however.  

\subsection{Binary effects}

Stripping of the envelopes of large stars could also be mediated by a
binary companion as the expanding giant overfills its Roche lobe
\citep{bailyn94}.  While this process should preferentially
occur in the centres of clusters where binaries will congregate as
a result of mass segregation, two-body relaxation progresses more
slowly in the most massive clusters.  Binaries should therefore sink 
to the cluster core more quickly in the least massive clusters,
contributing to an increase in the core binary fraction at a rate that
decreases with increasing cluster mass.  Observational evidence has
been found in support of this, most notably by \citet{sollima07} and
\citet{milone08} who found an anti-correlation between the cluster 
mass and the core binary fraction.  Any mechanism for RGB
depletion that relies on binary stars should therefore operate
more efficiently in the least massive cores where the binary
fraction is expected to be the highest.  Our results are consistent
with a surplus of giants in the least massive cores,
however.  This therefore argues against a binary mass-transfer origin
for RGB depletion in massive GC cores.  For similar reasons, it seems
unlikely that our result can be explained by collisions between RGB
stars and binaries.  If, on the other hand, 
RGB stars are more commonly found in binaries than are MS stars,
perhaps as a result of their larger cross-sections for tidal capture, 
binary stars could still be contributing to the observed trend.  Note
that in the cluster outskirts where the velocity dispersion has
dropped considerably from its central value, individual encounters are
more likely to result in tidal capture.  Since mass segregation should deliver
binaries to the core faster in the least massive clusters, a
larger fraction of their RGB stars could have hitched
a ride to the core as a binary companion.   However, both 
the average half-mass relaxation time of MW GCs and the RGB lifetime
tend to be on the order of a Gyr 
\citep{harris96, iben91}.  This does not leave much time for giants to be
captured into binaries and subsequently fall into the core before
evolving away from the RGB.

\subsection{Core helium-burning stars}

The fit for the HB sample is consistent with those of the RGB and MSTO
samples at the 3-sigma level so that we are unable to draw any
reliable conclusions for this stellar population.  The high
uncertainty stems from a number of outlying clusters.  
Selection effects and contamination from the Galaxy are likely to be
playing a role in this, in addition to our formulaic selection
criterion which may not be as suitable to the varying morphology of
the HB as it is to other stellar populations.  That is, the creation of
a purely photometry-based cluster-independent selection criterion may
not be possible for HB stars.  Given that stellar evolution effects
are expected to be the 
most dramatic at the end of the RGB lifetime, an interplay with the
cluster dynamics could also be contributing.  In particular, if the
central relaxation time is shorter than the HB lifetime, 
significant numbers of HB stars could be ejected from the core via
dynamical encounters as a result of having lost around a quarter of
their mass upon evolving off the tip of the RGB.  Moreover, since
stars expand considerably as they ascend the RGB, many of
the dynamical arguments presented in Section~\ref{dynamics} may more
strongly affect the size of the HB sample if they are the direct
evolutionary descendants of RGB stars.  Since at most a handful of
studies have been performed comparing the radial HB and RGB
distributions in GCs \citep[e.g.][]{iannicola09}, more data is needed
before any firm constraints can be placed on the source of the 
poor fit found for the HB sample.

\subsection{An evolutionary link with blue stragglers?}

The addition of a small number of extra RGB stars to every cluster is one
way to account for the observed sub-linear dependence on core
mass since the fractional increase in the size of
the RGB population will be substantially larger in the least massive
cores.  In log-log space, the result is a reduction of the
slope.  Since blue stragglers will evolve
into RGB and eventually HB stars \citep{sills09}, evolved BSs could be
the cause of a surplus of RGB (and possibly  
core helium-burning) stars in these clusters.  This scenario also
predicts that MSTO stars should scale slightly more linearly with core
mass since there should be a smaller contribution from evolved BSs, as
we have shown.  Given the fits for the RGB and BS  
samples presented in Section~\ref{results} and their corresponding
uncertainties, we find that the addition of evolved BSs to the RGB
populations could inflate the slope enough that the dependence on core
mass becomes linear.  Upon subtracting the BS sample from the RGB
sample, we find that the new fit is consistent with being linear:
\begin{equation}
\log (N_{RGB}) = (0.94 \pm 0.04)\log (N_{core}/10^3) + (1.97 \pm 0.02)
\end{equation}
The slope becomes larger if we have under-estimated the
number of BSs, perhaps as a result of selection effects, our adopted
selection criterion or a larger population size having existed in the
past.

\section{Summary} \label{summary}

In this paper, we have performed a cluster-to-cluster comparison
between the number of core RGB, MSTO \& HB stars 
and the total core mass.  We have introduced a technique for comparing
stellar populations in clusters that is well suited to studies of both
cluster and stellar evolution, in addition to the interplay thereof.
Using a sample of 56 GCs taken from Piotto et al.'s 2002 HST database,
we find a sub-linear scaling for RGB stars at the 3-$\sigma$ level,
whereas the relation is linear for MSTO stars.  While the preferential
self-enrichment of massive GCs, two-body relaxation, and evolved BSs
could all be contributing to the observed sub-linear dependence,
further studies with an emphasis on selection effects are needed in
order to better constrain the source of this curious observational result.

\section*{Acknowledgments}

We would like to thank an anonymous referee for a 
number of helpful suggestions, as well as Bill Harris, 
David Chernoff, Barbara Lanzoni and Francesco Ferraro for useful
discussions.  This research has been supported by NSERC as well as the
National Science Foundation under Grant No. PHY05-51164 to the Kavli
Institute for Theoretical Physics.

\bsp

\label{lastpage}

\end{document}